\documentclass[journal,twoside,12ptl]{IEEEtran}
\usepackage{fancyhdr}

\usepackage[ruled, lined, linesnumbered, commentsnumbered]{algorithm2e}
\usepackage{amsmath, amsthm, amssymb}
\usepackage{amsfonts}
\usepackage[dvips]{graphicx}
\usepackage[normalem]{ulem}
\usepackage{extarrows}
\usepackage{balance}
\usepackage{epsfig}
\usepackage{color}
\usepackage{bm}
\usepackage{enumerate}
\usepackage{latexsym}
\usepackage{graphics}
\usepackage{graphicx}
\usepackage{subfigure}
\usepackage{cite}
\usepackage{mathrsfs}
\usepackage{algorithmic}
\usepackage{stfloats}
\usepackage{extarrows}
\usepackage{upgreek}
\usepackage{caption}
\usepackage{epstopdf}
\begin{document}

\title{RIS-Assisted Cooperative NOMA with SWIPT}
\author{
Juanjuan Ren, Xianfu~Lei, Zhangjie Peng, Xiaohu Tang, and Octavia A. Dobre

\thanks{J. Ren, X. Lei and X. Tang are with the School of Information Science and Technology, Southwest Jiaotong University, Chengdu 610031, China (e-mail: juanjuanren@foxmail.com; xflei@swjtu.edu.cn; xhutang@swjtu.edu.cn).}
\thanks{Z. Peng is with the College of Information, Mechanical, and Electrical Engineering, Shanghai Normal University, Shanghai 200234,
China (email: pengzhangjie@shnu.edu.cn).}
\thanks{O. A. Dobre is with the Faculty of Engineering and Applied Science, Memorial University, St. John's, NL A1B 3X9, Canada (e-mail: odobre@mun.ca).}}

\maketitle
\vspace{-0.5 in}
\begin{abstract}
This paper studies the application of reconfigurable intelligent surface (RIS) to cooperative non-orthogonal multiple access (C-NOMA) networks with simultaneous wireless information and power transfer (SWIPT). We aim for maximizing the rate of the strong user with guaranteed weak user's quality of service (QoS) by jointly optimizing power splitting factors, beamforming coefficients, and RIS reflection coefficients in two transmission phases. The formulated problem is difficult to solve due to its complex and non-convex constraints. To tackle this challenging problem, we first use alternating optimization (AO) framework to transform it into three subproblems, and then use the penalty-based arithmetic-geometric mean approximation (PBAGM) algorithm and the successive convex approximation (SCA)-based method to solve them. Numerical results verify the superiority of the proposed algorithm over the baseline schemes.

\end{abstract}


\section{Introduction}
\IEEEPARstart{R}{ECENTLY}, reconfigurable intelligent surface (RIS) has been widely studied as an emerging technology\cite{wu2019towards,guo2020weighted,zhou2020robust}. A RIS consists of a large number of passive elements, which can reconfigure the wireless propagation channel between transceivers by adjusting the phase of each passive element on the surface. In particular, compared with the conventional relay and multiple-input multiple-output (MIMO) techniques, RIS can achieve signal enhancement and interference suppression in a cost-effective and energy-efficient manner \cite{wu2021intelligent}. Besides, RISs can be flexibly deployed in existing communication systems. The above-mentioned benefits of RIS have motivated an upsurge of interest in the integration of RISs in various scenarios \cite{bai2021resource,mao2021intelligent}.

In order to improve the communication reliability for the cell-edge users, cooperative non-orthogonal multiple access (C-NOMA) has been studied in  \cite{ding2015cooperative,li2018cooperative,cao2020power}. The C-NOMA strategy was first proposed in \cite{ding2015cooperative}, in which the strong user is used as a relay to help the transmission of the weak user. The achievable rate maximization problem in C-NOMA system with MIMO channels was studied in \cite{li2018cooperative}, while \cite{cao2020power} investigated the power optimization for enhancing secrecy of C-NOMA networks. In addition, simultaneous wireless information and power transfer (SWIPT) has been envisioned as a promising technology to reduce the cost of battery replacement and provide perpetual energy supply. Thus, in order to alleviate the energy constraint, the application of SWIPT to C-NOMA system has also been studied in some prior works, such as \cite{liu2016cooperative} and \cite{xu2017joint}. Liu \emph{et al.} considered a C-NOMA with SWPIT network, in which strong users act as energy harvesting relays to help weak users, and derived the outage probability and system throughput in \cite{liu2016cooperative}. By jointly optimizing beamforming vectors and power splitting factor, a problem of maximizing the achievable rate of strong users while guaranteeing the quality of service (QoS) of weak users was studied in \cite{xu2017joint}.

To the best of our knowledge, this is the first work to investigate the application of RIS to SWIPT C-NOMA networks. Specifically, an RIS-assisted two-phase transmission protocol is designed and an optimization problem is formulated, which maximizes the rate of the strong user with better channel quality while satisfying the week user's QoS requirement. This problem is a multivariate coupling problem with complex constraints about the sum of fractional and linear functions with respect to the RIS phase shift, which was never studied in the published literature. In order to solve this challenging problem, we propose an iterative algorithm based on the alternate optimization (AO) framework, penalty-based arithmetic-geometric mean approximation (PBAGM) and successive convex approximation (SCA)-based method.
\section{System Model}
\begin{figure}[t!]
   \centering
   \includegraphics[width=1\linewidth]{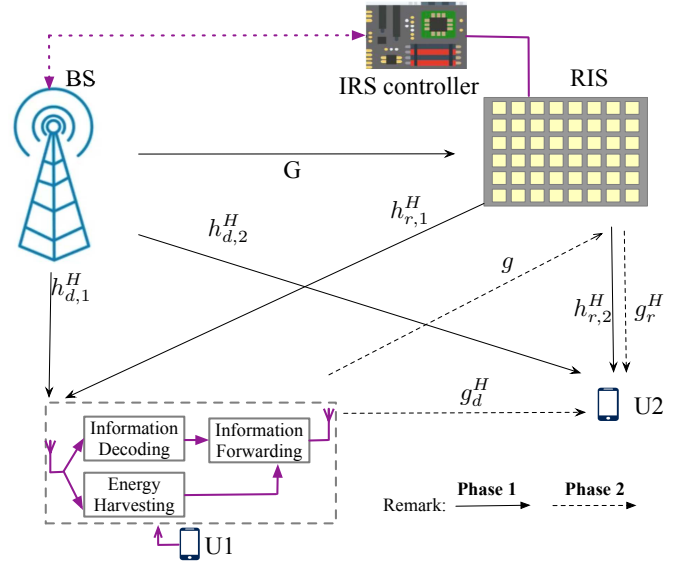}
   \captionsetup{font={normal}}
    \caption{RIS-assisted C-NOMA with SWIPT.}
   \label{fig:example}
   \vspace{-0.2 in}
\end{figure}
As shown in Fig. 1, we consider an RIS-assisted C-NOMA with SWIPT network, which consists of one AP equipped with $N$ antennas, an RIS with $M$ reflection elements and two single-antenna users ($U_1$ and $U_2$). It is assumed that $U_1$ (strong user) with a better channel condition than $U_2$ (weak user), and $U_1$ can assist the AP to transmit signal to $U_2$. Moreover, $U_1$ is energy constrained, which needs to harvest radio-frequency (RF) energy from the AP to power its relaying operations, and power splitting (PS) scheme is employed to perform SWIPT.

We assume that the AP and all the other devices operate over the same frequency band, i.e., channel reciprocity holds for all channels. Moreover, it is assumed that all channels are subject to flat fading, and the AP can acquire perfect channel state information (CSI)\cite{wu2019towards}. The channel coefficients from the AP to $U_i$ ($i\in{1,2}$), from the AP to the RIS, and from the RIS to $U_i$ are denoted by $\mathbf{h}_{d,i}\in \mathbb{C}^{N \times1}$, $\mathbf{G} \in \mathbb{C}^{M \times N}$, and $\mathbf{h}_{r,i}\in \mathbb{C}^{M \times1}$, respectively. Moreover, the channel coefficients from $U_1$ to $U_2$, from $U_1$ to the RIS, and from the RIS to $U_2$ are denoted by $g_d\in \mathbb{C}$, $\mathbf{g}\in \mathbb{C}^{M \times1}$, and $\mathbf{g}_{r}\in \mathbb{C}^{M \times1}$, respectively.

\subsection{Transmission Protocol}
For the above described network, the transmission process is divided into two phases. In phase $1$, the AP transmits the combined signal to $U_1$ and $U_2$ by applying superposition coding with the assist of RIS. The received signal at $U_1$ is divided into two parts, one for energy harvesting (EH) and the other for information decoding. In phase $2$, $U_1$ forwards the information to $U_2$ by using the harvested energy, meanwhile, the signal is reflected to $U_2$ by the RIS. Moreover, the two phases are assumed to have the same transmission duration $\tau$ and the two-phase transmission protocol is described as follows:

$\bf{Phase\,\,1}$\textbf{: Direct Transmission.} During this phase, the AP transmits signal ${\bf{x}}={\bf{w}}_1x_1+{\bf{w}}_2x_2$ to both users, where ${\bf{w}}_i (i\in\{1,2\})$ stands for the corresponding transmit beamforming vectors and $x_i (i\in\{1,2\})$ is the transmitted symbol intended for $U_i$, with $\mathbb{E}\{|x_i|^2\}=1$. The received signal at $U_i$ can be expressed as:
\begin{equation}
y^{(1)}_i=(\mathbf{h}^H_{r,i} \mathbf{\Theta}_1 \mathbf{G}+\mathbf{h}^H_{d,i})\mathbf{x}+ n_i,
  \end{equation}
  where $n_i \sim \mathcal{CN}(0,\sigma_i^2)$ is the additive white Gaussian noise (AWGN) and $\sigma_i^2$ stands for the noise variance at $U_i$. Moreover, the diagonal matrix $\mathbf{\Theta}_{1}=\mathrm{diag}(\theta_{1,1},\theta_{1,2},...,\theta_{1,M}) \in \mathbb{C}^{M \times M}$ represents the reflection phase-shift matrix in phase 1 and $\theta_{1,m}=e^{j\Phi_{1,m}}$ $\left(\Phi_{1,m}\in [0, 2\pi ]\right)$. Letting $\beta \in [0,1]$ to denote the power splitting factor of $U_1$ used for EH. Combined with the assumption that the transmission duration
$\tau=1/2$, the harvested energy at $U_1$ can be given as
   \begin{align}
    E=&\frac{1}{2}\beta\eta\left(|(\mathbf{h}^H_{r,1} \mathbf{\Theta}_1 \mathbf{G}+\mathbf{h}^H_{d,1})\mathbf{w}_1|^2+\right.\nonumber\\
   &\left. |(\mathbf{h}^H_{r,1} \mathbf{\Theta}_1 \mathbf{G}+\mathbf{h}^H_{d,1})\mathbf{w}_2|^2\right),
    \end{align}
 where $\eta\in(0,1]$ is the energy conversion efficiency. According to the successive interference cancellation (SIC) principle of NOMA, the strong user $U_1$ first decodes the information of the weak user $U_2$, and the corresponding SINR for $U_1$ to decode the information of $U_2$ can be given by
     \begin{equation}
  \mathrm{SINR}^{(1)}_{1\to 2}=\frac{(1-\beta)|(\mathbf{h}^H_{r,1} \mathbf{\Theta}_1 \mathbf{G}+\mathbf{h}^H_{d,1})\mathbf{w}_2|^2}{(1-\beta)|(\mathbf{h}^H_{r,1} \mathbf{\Theta}_1 \mathbf{G}+\mathbf{h}^H_{d,1})\mathbf{w}_1|^2+ \sigma^2_1}.
  \end{equation}
  $U_1$  subtracts the information of $U_2$ from the combined signal. Thus, the SNR of $U_1$ for decoding its own information can be expressed as
   \begin{equation}
   \mathrm{SNR}^{(1)}_{1} = (1-\beta)\frac{|(\mathbf{h}^H_{r,1} \mathbf{\Theta}_1 \mathbf{G}+\mathbf{h}^H_{d,1})\mathbf{w}_1|^2}{\sigma^2_1}.
   \end{equation}
  The SINR of $U_2$ can be described as
   \begin{equation}
  \mathrm{SINR}^{(1)}_{2}=\frac{|(\mathbf{h}^H_{r,2} \mathbf{\Theta}_1 \mathbf{G}+\mathbf{h}^H_{d,2})\mathbf{w}_2|^2}{|(\mathbf{h}^H_{r,2} \mathbf{\Theta}_1 \mathbf{G}+\mathbf{h}^H_{d,2})\mathbf{w}_1|^2+ \sigma^2_2}.\label{phase12}
  \end{equation}

 $\bf{Phase\,\,2}$\textbf{: Cooperative Transmission.} During this phase, $U_1$ transmits signal $x_2$ to $U_2$ with the assist of RIS by using the harvested energy in phase 1. The received signal of $U_2$ is
   \begin{equation}
y^{(2)}_2=\sqrt{P_t}(g^H_d+g^H_r \mathbf{\Theta}_2 g)x_2+n_2,
  \end{equation}
where the diagonal matrix $\mathbf{\Theta}_{2}=\mathrm{diag}(\theta_{2,1},\theta_{2,2},...,\theta_{2,M}) \in \mathbb{C}^{M \times M}$ is the reflection phase-shift matrix in this phase and $\theta_{2,m}=e^{j\Phi_{2,m}}$ with  $\Phi_{2,m}\in [0, 2\pi ]$, moreover, $P_t$ stands for the transmitted power of $U_1$ and it can be expressed as
  \begin{align}
 P_t=E/(1/2)&=\beta \eta \left(|(\mathbf{h}^H_{r,1} \mathbf{\Theta}_1 \mathbf{G}+\mathbf{h}^H_{d,1})\mathbf{w}_1|^2+\right.\nonumber\\
 &\quad\left. (\mathbf{h}^H_{r,1} \mathbf{\Theta}_1 \mathbf{G}+\mathbf{h}^H_{d,1})\mathbf{w}_2|^2\right).
 \end{align}
 Thus, the SNR of $U_2$ in this phase is given by
 \begin{align}
  \mathrm{SNR}^{(2)}_{2}= &\beta \eta |g^H_d+g^H_r \mathbf{\Theta}_2 g|^2 \left(|(\mathbf{h}^H_{r,1} \mathbf{\Theta}_1 \mathbf{G}+\mathbf{h}^H_{d,1})\mathbf{w}_1|^2 \right.\nonumber\\
& \left.+|(\mathbf{h}^H_{r,1} \mathbf{\Theta}_1 \mathbf{G}+\mathbf{h}^H_{d,1})\mathbf{w}_2|^2\right)/\sigma^2_2.\label{phase22}
  \end{align}
$U_2$ combined its received signals in phase 1 and phase 2 to decoding $x_2$ by employing maximal-ratio combining (MRC) \cite{xu2017joint}, i.e., the corresponding SINR can be written as
   \begin{align}
  \mathrm{SINR}_{2}=&\mathrm{SINR}^{(1)}_{2}+ \mathrm{SNR}^{(2)}_{2} \nonumber\\
  = &\frac{|(\mathbf{h}^H_{r,1} \mathbf{\Theta}_1 \mathbf{G}+\mathbf{h}^H_{d,1})\mathbf{w}_2|^2}{|(\mathbf{h}^H_{r,1} \mathbf{\Theta}_1 \mathbf{G}+\mathbf{h}^H_{d,1})\mathbf{w}_1|^2+ \sigma^2_1}+\nonumber\\
 & \beta \eta |g^H_d+g^H_r \mathbf{\Theta}_2 g|^2 \left(|(\mathbf{h}^H_{r,1} \mathbf{\Theta}_1 \mathbf{G}+\mathbf{h}^H_{d,1})\mathbf{w}_1|^2 \right.\nonumber\\
& \left.+|(\mathbf{h}^H_{r,1} \mathbf{\Theta}_1 \mathbf{G}+\mathbf{h}^H_{d,1})\mathbf{w}_2|^2\right)/\sigma^2_2.\label{phase22}
 \end{align}
 \subsection{Problem Formulation}
In this section, we aim at maximizing the achievable rate of $U_1$ while guaranteeing the QoS of $U_2$. The joint design of the PS factor at $U_1$, the transmit beamforming at the AP, as well as the RIS reflection coefficients in both phases can be mathematically formulated as
  \begin{align}
\text{\bf{P1}}:&\underset{\substack{\beta, {\bf{w}}_1, {\bf{w}}_2, \\{\bf{\Theta}}_1,{\bf{\Theta}}_2}}{\text{\text{max}}} \frac{1}{2}\log_2\left(1+(1-\beta)\frac{|(\mathbf{h}^H_{r,1} \mathbf{\Theta}_1 \mathbf{G}+\mathbf{h}^H_{d,1})\mathbf{w}_1|^2}{\sigma^2_1}\right)\nonumber\\
&\quad\text{\text{s.t.}}\,\,\,\,\mathrm{C}_1:\frac{1}{2}\log_2\left(1+\right.\nonumber\\
&\quad\quad\quad\quad\left.{\frac{(1-\beta)|(\mathbf{h}^H_{r,1} \mathbf{\Theta}_1 \mathbf{G}+\mathbf{h}^H_{d,1})\mathbf{w}_2|^2}{(1-\beta)|(\mathbf{h}^H_{r,1} \mathbf{\Theta}_1 \mathbf{G}+\mathbf{h}^H_{d,1})\mathbf{w}_1|^2+ \sigma^2_1}}\right)\geq \gamma_2, \nonumber\\
&\quad\quad\,\,\,\,\,\mathrm{C}_2:\frac{1}{2}\log_2\left(1+\frac{|(\mathbf{h}^H_{r,2} \mathbf{\Theta}_1 \mathbf{G}+\mathbf{h}^H_{d,2})\mathbf{w}_2|^2}{|(\mathbf{h}^H_{r,2} \mathbf{\Theta}_1 \mathbf{G}+\mathbf{h}^H_{d,2})\mathbf{w}_1|^2+ \sigma^2_2}+\right.\nonumber\\
&\quad\quad\quad\quad\quad \beta \eta |g^H_d+g^H_r \Theta_2 g|^2 \left(|(\mathbf{h}^H_{r,1} \mathbf{\Theta}_1 \mathbf{G}+\mathbf{h}^H_{d,1})\mathbf{w}_1|^2 \right.\nonumber\\
&\quad\quad\quad\quad\quad \left.\left.+|(\mathbf{h}^H_{r,1} \mathbf{\Theta}_1 \mathbf{G}+\mathbf{h}^H_{d,1})\mathbf{w}_2|^2\right)/\sigma^2_2\right)\geq \gamma_2, \nonumber\\
&\quad\quad\,\,\,\,\,\mathrm{C}_3: {||\mathbf{w}_1||^2+||\mathbf{w}_2||^2\leq P_s}, \nonumber\\
&\quad\quad\,\,\,\,\,\mathrm{C}_4:{0\leq \beta \leq 1}, \nonumber\\
&\quad\quad\,\,\,\,\,\mathrm{C}_5:{0\leq \Phi_{1,m} \leq 2\pi}, \forall m, \nonumber\\
&\quad\quad\,\,\,\,\,\mathrm{C}_6:{0\leq \Phi_{2,m} \leq 2\pi}, \forall m,
 \label{optun}
 \end{align}
where $\gamma_2$ is the target rate of $U_2$. The objective function is the achievable rate of $U_1$. The constraint $\mathrm{C}_1$ guarantees that $U_1$ can successfully decode $x_2$. The constraint $\mathrm{C}_2$ is imposed to ensure the QoS requirement of $U_1$. Moreover, the constraints of power budget at the AP and the power splitting factor are characterized by $\mathrm{C}_3$ and $\mathrm{C}_4$, respectively. Finally, $\mathrm{C}_5$ and $\mathrm{C}_6$ represent the passive RIS phase shift constraints in phase 1 and phase 2, respectively.
\subsection{The Proposed Algorithm}
Problem $\text{\bf{P1}}$ is a multiple-variable non-convex optimization problem, the main challenges to solve which can be summarized as follows: i) the PS factor, beamforming coefficients and  RIS reflection matrix are coupled both in the objective function and the constraints $\mathrm{C}_1$ and $\mathrm{C}_2$; ii) the non-convex unit-modulus constraints $\mathrm{C}_5$ and $\mathrm{C}_6$; iii) the left hand side of the non-convex QoS constraint $\mathrm{C}_2$ is the sum of a fraction and a linear function. In general, it is difficult to directly solve this problem. In the following, we first apply AO method to decouple $\text{\bf{P1}}$ to three sub-problems, and then tackle each sub-problem separately.

Firstly, optimize $\mathbf{w}_1$, $\mathbf{w}_2$, and $\beta$ with given RIS reflection coefficients $\mathbf{\Theta}_1$ and $\mathbf{\Theta}_2$. For the ease of representation, let $\mathbf{\theta}_1=[\theta_{1,1},\theta_{1,2},...,\theta_{1,M}]^H \in \mathbb{C}^{M \times1}$ and $\mathbf{\tilde{G}}_{r,i}=\mathrm{diag}(\mathbf{h}^H_{r,i})\mathbf{G}$. Then we have
 \begin{align}
 (\mathbf{h}^H_{r,i} \mathbf{\Theta}_1 \mathbf{G}+\mathbf{h}^H_{d,i})\mathbf{w}_i|^2&=| (\mathbf{\theta}^H_1\mathbf{\tilde{G}}_{r,i}+\mathbf{h}^H_{d,i})\mathbf{w}_i|^2\nonumber\\
& \xlongequal{ \mathbf{\tilde{h}}^H_i=\mathbf{\theta}^H_1\mathbf{\tilde{G}}_{r,i}+\mathbf{h}^H_{d,i}}|\mathbf{\tilde{h}}^H_i\mathbf{w}_i|^2\nonumber\\
&\xlongequal[\mathbf{H}_i=\mathbf{\tilde{h}}_i\mathbf{\tilde{h}}^H_i]{\mathbf{W}_i=\mathbf{w}_i\mathbf{w}^H_i}\mathrm{Tr}(\mathbf{H}_i\mathbf{W}_i).
\end{align}
 Then, by letting $g^H_d+g^H_r \Theta_2 g= \tilde{g}$, we can obtain the following subproblem
  \begin{align}
\text{\bf{P2}}:&\underset{\beta, {\bf{W}}_1, {\bf{W}}_2}{\text{\text{max}}}\frac{1}{2}\log_2\left(1+\frac{(1-\beta)\mathrm{Tr}(\mathbf{H}_1\mathbf{W}_1)}{\sigma^2_1}\right)\nonumber\\
&\quad\text{\text{s.t.}}\,\,\,\,\mathrm{C}_1:{\frac{1}{2}\log_2(1\!+\!\frac{(1-\beta)\mathrm{Tr}(\mathbf{H}_1\mathbf{W}_2)}{(1-\beta)\mathrm{Tr}(\mathbf{H}_1\mathbf{W}_1)+ \sigma^2_1}})\geq \gamma_2, \nonumber\\
&\quad\quad\,\,\,\,\,\mathrm{C}_2:\frac{1}{2}\log_2\left(1+\frac{\mathrm{Tr}(\mathbf{H}_2\mathbf{W}_2)}{\mathrm{Tr}(\mathbf{H}_2\mathbf{W}_1)+ \sigma^2_2}+\right.\nonumber\\
&\quad\quad\quad\quad\quad \left.\beta \eta |{\tilde{g}}|^2 \mathrm{Tr}(\mathbf{H}_1\left(\mathbf{W}_1+\mathbf{W}_2)\right)/\sigma^2_2\right) \geq \gamma_2, \nonumber\\
&\quad\quad\,\,\,\,\,\mathrm{C}_3: {\mathrm{Tr}(\mathbf{W}_1)+\mathrm{Tr}(\mathbf{W}_2)\leq P_s}, \nonumber\\
&\quad\quad\,\,\,\,\,\mathrm{C}_4:{0\leq \beta \leq 1}, \nonumber\\
&\quad\quad\,\,\,\,\,\mathrm{C}_5:{\mathrm{rank}\left(\mathbf{W}_i\right)\leq1}, i \in\{1,2\}.
 \end{align}
Problem \text{\bf{P2}} jointly optimizes beamforming and power-splitting, which can be solved by using the proposed algorithm in \cite{xu2017joint}.

Secondly, optimize $\Theta_1$ with given $\beta$, $\mathbf{W}_1$, $\mathbf{W}_2$ and $\Theta_2$. Let $\mathbf{\tilde{G}}\mathbf{w}_1=\mathbf{a}_1$, $\mathbf{h}^H_{d,1}\mathbf{w}_1={b}_1$, and introducing auxiliary variable $\mathbf{\tilde{\theta}}_1={\left[
 \begin{matrix}
  \mathbf{\theta}_1\\
1\\
      \end{matrix}
  \right]}$, one can get
   \begin{align}
    (\mathbf{h}^H_{r,1} \mathbf{\Theta}_1 \mathbf{G}+\mathbf{h}^H_{d,1})\mathbf{w}_1|^2=| (\mathbf{\theta}^H_1\mathbf{\tilde{G}}_{r,1}+\mathbf{h}^H_{d,1})\mathbf{w}_1|^2 \nonumber\\
 =\underbrace{\mathbf{\theta}^H_1\mathbf{a}_1\mathbf{a}^H_1\mathbf{\theta}_1+\mathbf{\theta}^H_1\mathbf{a}_1{b}^H_1+ {b}_1\mathbf{a}^H_1\mathbf{\theta}_1}_{f(\mathbf{\theta}_1,\mathbf{a}_1, {b}_1)}+|b_1|^2,
 \end{align}
  where $f(\mathbf{\theta}_1,\mathbf{a}_1, {b}_1)$ can be rewritten as
   \begin{align}
  f(\mathbf{\theta}_1,\mathbf{a}_1, {b}_1)&= \underbrace{[\mathbf{\theta}^H_1\quad1]}_{\mathbf{\tilde{\theta}}^H_1}  \underbrace{\left[
 \begin{matrix}
  \mathbf{a}_1\mathbf{a}^H_1&  \mathbf{a}_1 {b}^H_1\\
{b}_1\mathbf{a}^H_1& 0\\
      \end{matrix}
  \right]}_{ \mathbf{R}_1} \underbrace{\left[
 \begin{matrix}
  \mathbf{\theta}_1\\
1\\
      \end{matrix}
  \right]}_{\mathbf{\tilde{\theta}}_1}\nonumber\\
 & =\mathbf{\tilde{\theta}}^H_1\mathbf{R}_1\mathbf{\tilde{\theta}}_1 \xlongequal{ \mathbf{\tilde{\Theta}}_1=\mathbf{\tilde{\theta}}_1\mathbf{\tilde{\theta}}^H_1}\mathrm{Tr}\left(\mathbf{R}_1\mathbf{\tilde{\Theta}}_1\right).\label{14}
  \end{align}
 Similarly, we can get the following equations
    \begin{subequations}
\begin{equation}
|(\mathbf{h}^H_{r,1} \mathbf{\Theta}_1 \mathbf{G}+\mathbf{h}^H_{d,1})\mathbf{w}_2|^2= \mathrm{Tr}\left(\mathbf{R}_2\mathbf{\tilde{\Theta}}_1\right)+|b_2|^2,
\end{equation}
\begin{equation}
|(\mathbf{h}^H_{r,2} \mathbf{\Theta}_1 \mathbf{G}+\mathbf{h}^H_{d,2})\mathbf{w}_1|^2= \mathrm{Tr}\left(\mathbf{R}_3\mathbf{\tilde{\Theta}}_1\right)+|b_3|^2,
\end{equation}
\begin{equation}
|(\mathbf{h}^H_{r,2} \mathbf{\Theta}_1 \mathbf{G}+\mathbf{h}^H_{d,2})\mathbf{w}_2|^2= \mathrm{Tr}\left(\mathbf{R}_4\mathbf{\tilde{\Theta}}_1\right)+|b_4|^2.
\end{equation}
\label{15}
\end{subequations}
Moreover, the channel gain in phase 2  can be repressed as
     \begin{align}
|g^H_d+g^H_r\mathbf{ \Theta}_2 g|^2&=|g^H_d+\mathbf{\theta}^H_2\tilde{g}|^2\nonumber\\
 &=\underbrace{\mathbf{\theta}^H_2\tilde{g}\tilde{g}^H\mathbf{\theta}_2+\mathbf{\theta}^H_2\tilde{g}g_d+g^H_d\tilde{g}^H\mathbf{\theta}_2}_{g(\mathbf{\theta}_2,g,\tilde{g})}+|\tilde{g}_d^H|^2,
 \end{align}
  where $g(\mathbf{\theta}_2,g,\tilde{g})$ can be rewritten as
   \begin{align}
 g(\mathbf{\theta}_2,g,\tilde{g})&= \underbrace{[\mathbf{\theta}^H_2\quad1]}_{\mathbf{\tilde{\theta}}^H_2}  \underbrace{\left[
 \begin{matrix}
 \tilde{g}\tilde{g}^H&  \tilde{g}g_d \\
g^H_d\tilde{g}^H
& 0\\
      \end{matrix}
  \right]}_{ \mathbf{R}_5} \underbrace{\left[
 \begin{matrix}
  \mathbf{\theta}_2\\
1\\
      \end{matrix}
  \right]}_{\mathbf{\tilde{\theta}}_2}\nonumber\\
 & =\mathbf{\tilde{\theta}}^H_2\mathbf{R}_5\mathbf{\tilde{\theta}}_2 \xlongequal{ \mathbf{\tilde{\Theta}}_2=\mathbf{\tilde{\theta}}^H_2\mathbf{\tilde{\theta}}_2}\mathrm{Tr}\left(\mathbf{R}_5\mathbf{\tilde{\Theta}}_2\right).\label{17}
         \end{align}
By substituting (\ref{14}), (\ref{15}) and (\ref{17}) into $\text{\bf{P1}}$, $\text{\bf{P1}}$ can be rewritten as
 \begin{align}
\text{\bf{P3}}: &\underset{\mathbf{\tilde{\Theta}}_1}{\text{\text{max}}} \quad \frac{1}{2}\log_2\left(1+\frac{(1-\beta)\left(\mathrm{Tr}\left(\mathbf{R}_1\mathbf{\tilde{\Theta}}_1\right)+ |\mathbf{b}_1|^2\right)}{\sigma^2\!_1}\right)\nonumber\\
&\quad\text{\text{s.t.}}\,\,\,\,\mathrm{C}_1: (1\!-\!\beta)\left(\mathrm{Tr}\left(\mathbf{R}_2\mathbf{\tilde{\Theta}}_1\right) \!+\! |\mathbf{b}_2|^2\right) \geq \left(2^{2 {\gamma_2}}\!-\!1\right)\nonumber\\
&\quad\quad\quad\quad\quad\left((1-\beta)\left(\mathrm{Tr}\left(\mathbf{R}_1\mathbf{\tilde{\Theta}}_1\right) +|\mathbf{b}_1|^2\right)+\sigma^2_1\right),\nonumber\\
&\quad\quad\,\,\,\,\,\mathrm{C}_2: \frac{\mathrm{Tr}\left(\mathbf{R}_4\mathbf{\tilde{\Theta}}_1\right) + |\mathbf{b}_4|^2}{\mathrm{Tr}\left(\mathbf{R}_3\mathbf{\tilde{\Theta}}_1\right) + |\mathbf{b}_3|^2+\sigma^2_2}+\beta\eta\times\nonumber\\
&\quad\quad\quad\quad\left(\mathrm{Tr}\left(\mathbf{R}_5\mathbf{\tilde{\Theta}}_2\right) \!+\! |\mathbf{b}_5|^2\right)\left(\mathrm{Tr}\left(\left(\mathbf{R}_1+\mathbf{R}_2\right)\mathbf{\tilde{\Theta}}_1\right)\right.\nonumber\\
&\quad\quad\quad\quad\quad\quad\quad\quad\left.+ |\mathbf{b}_1|^2+ |\mathbf{b}_2|^2\right)/\sigma^2_2 \geq 2^{2\gamma_2}-1, \nonumber\\
&\quad\quad\,\,\,\,\,\mathrm{C}_3:{{\mathrm{rank}\left(\mathbf{\tilde{\Theta}}_1\right)=1}}, \nonumber\\
&\quad\quad\,\,\,\,\,\mathrm{C}_4:{\left(\mathbf{\tilde{\Theta}}_1\right)_{m,m}=1}, m=1,2,...,M+1.\label{theta_1}
 \end{align}
It can be observed that \text{\bf{P3}} is difficult to solve directly, due to the non-convex constraints $\mathrm{C}_2$ and $\mathrm{C}_3$. Note that $\mathrm{C}_2$ is the sum of fraction and linear function with respect to $\mathbf{\tilde{\Theta}}_1$. We introduce an auxiliary variable $\mathcal{X}$ to replace the fractional function term. Then, \text{\bf{P3}} can be equivalently transformed into
 \begin{align}
\text{\bf{P4}}:&\underset{\mathbf{\tilde{\Theta}}_1,\mathcal{X} }{\text{\text{max}}} \frac{1}{2}\log_2\left(1+\frac{(1-\beta)\left(\mathrm{Tr}\left(\mathbf{R}_1\mathbf{\tilde{\Theta}}_1\right)+ |\mathbf{b}_1|^2\right)}{\sigma^2_1}\right)\nonumber\\
&\quad\text{\text{s.t.}}\,\,\,\,\mathrm{C}_1: (1\!-\!\beta)\left(\mathrm{Tr}\left(\!\mathbf{R}_2\mathbf{\tilde{\Theta}}_1\!\right) \!+\! |\mathbf{b}_2|^2\right) \!\geq\! \left(2^{2\gamma_2}\!-\!1\right)\nonumber\\
&\quad\quad\quad\quad\quad\left((1-\beta)\left(\mathrm{Tr}\left(\mathbf{R}_1\mathbf{\tilde{\Theta}}_1\right) +|\mathbf{b}_1|^2\right)+\sigma^2_1\right),\nonumber\\
&\quad\quad\,\,\,\,\,\mathrm{C}_2: \frac{\mathrm{Tr}\left(\mathbf{R}_4\mathbf{\tilde{\Theta}}_1\right) + |\mathbf{b}_4|^2}{\mathrm{Tr}\left(\mathbf{R}_3\mathbf{\tilde{\Theta}}_1\right) + |\mathbf{b}_3|^2+\sigma^2_2}\geq\mathcal{X}\nonumber\\
&\quad\quad\,\,\,\,\,\mathrm{C}_3: \mathcal{X}+\beta\eta\left(\mathrm{Tr}\left(\mathbf{R}_5\mathbf{\tilde{\Theta}}_2\right) + |\mathbf{b}_5|^2\right)\times\nonumber\\
&\quad\quad\quad\quad\,\,\,\,\,\left(\mathrm{Tr}\left(\left(\mathbf{R}_1+\mathbf{R}_2\right)\mathbf{\tilde{\Theta}}_1\right)\!+\! |\mathbf{b}_1|^2\!+ \!|\mathbf{b}_2|^2\right)/\sigma^2_2 \nonumber\\
&\quad\quad\quad\quad\quad\quad\quad\quad\quad\quad\quad\quad\quad\quad\quad \geq 2^{2\gamma_2}-1, \nonumber\\
&\quad\quad\,\,\,\,\,\mathrm{C}_4:{{\mathrm{rank}\left(\mathbf{\tilde{\Theta}}_1\right)=1}}, \nonumber\\
&\quad\quad\,\,\,\,\,\mathrm{C}_5:{\left(\mathbf{\tilde{\Theta}}_1\right)_{m,m}=1}, m=1,2,...,M+1.\label{theta_11}
 \end{align}
 Although \text{\bf{P4}} is more tractable than \text{\bf{P3}}, it is still a non-convex problem due to $\mathrm{C}_2$ and $\mathrm{C}_4$. By using the arithmetic-geometric mean inequality, $\mathrm{C}_2$ can be approximated as
 \begin{align}
 \mathcal{X}\left(\mathrm{Tr}\left(\!\mathbf{R}_3\mathbf{\tilde{\Theta}}_1\!\right)\!+\!|b_3|^2\right)&\!\leq\!\frac{(y^{(n)} \mathcal{X})^2\!+\!\left(\!\frac{\left(\mathrm{Tr}\left(\mathbf{R}_3\mathbf{\tilde{\Theta}}_1\right)+|b_3|^2\right)}{y^{(n)}}\right)^2}{2}\nonumber\\
 & \leq\mathrm{Tr}\left(\mathbf{R}_4\mathbf{\tilde{\Theta}}_1\right)\!+\!|b_4|^2\!\!-\!\!\mathcal{X}\sigma^2_2,
 \end{align}
 where  $y^{(n)}$ can be updated by $  y^{(n)}=\frac{\mathcal{X}\left(\mathrm{Tr}\left(\mathbf{R}_3\mathbf{\tilde{\Theta}}_1\right)+|b_3|^2\right)}{\mathcal{X}}$.
For the non-convex rank one constraint $\mathrm{C}_4$, we first rewrite it as $\mathrm{Tr}\left(\mathbf{\tilde{\Theta}}_1\right)-||\mathbf{\tilde{\Theta}}_1||_2= 0$. However, it is still non-convex since the left hand side of it is the difference of two convex functions. Thus, we linearize the second term as $\langle\partial||\mathbf{\tilde{\Theta}}_1||^{(n)}_2,\mathbf{\tilde{\Theta}}_1\rangle$ and then use the penalty-based method to deal with it, i.e., in the $n^\mathrm{th}$ iteration, we need to solve the optimization problem $\text{\bf{P5}}$.
   \begin{algorithm}[htb]
{{
 \caption{Penalty-based arithmetic-geometric mean approximation (PBAGM) algorithm}
\label{alg:Framwork}
\begin{algorithmic}[1]
\STATE {\bf{Input:}} $c^{(0)}$, $y^{(0)}$;
\STATE{\bf{letting} $\Delta={(1-\beta)\left(\mathrm{Tr}\left(\mathbf{R}_1\mathbf{\tilde{\Theta}}^{(n)}_1\right)\right)\!-\!(1-\beta)\left(\mathrm{Tr}\left(\mathbf{R}_1\mathbf{\tilde{\Theta}}^{(n-1)}_1\right)\right)}$, $\delta=\mathrm{Tr}\left(\mathbf{\tilde{\Theta}}^{(n)}_1\right)-||\mathbf{\tilde{\Theta}}^{(n)}_1||_2$};
\STATE {\bf while} {$\Delta \geq eps$ and $\delta \geq eps$}\\
{\bf do}\\
\STATE Compute the optimal $\mathbf{\tilde{\Theta}}_1$ and $\mathcal{X}$ by solving P5;
\STATE Update the penalty parameter $c$ and the auxiliary variable $y$ according to $c^{(n+1)}=\rho c^{(n)}$ and $y^{(n)}= \frac{\mathcal{X}^{(n)}\left(\mathrm{Tr}\left(\mathbf{R}_3\mathbf{\tilde{\Theta}}^{(n)}_1\right)+|b_3|^2\right)}{\mathcal{X}^{(n)}}$ respectively.\\
{\bf end while}
\STATE Compute the RIS reflection coefficient in phase as $\mathbf{{\theta}}_1=\mathbf{\tilde{\Theta}}_1(1:M,M+1)$.
\end{algorithmic}}}
\end{algorithm}

  \begin{align}
\text{\bf{P5}}:&\underset{\mathbf{\tilde{\Theta}}_1,\mathcal{X} }{\text{\text{max}}} \quad \frac{1}{2}\log_2\left(1\!+\!\frac{(1-\beta)\left(\mathrm{Tr}\left(\mathbf{R}_1\mathbf{\tilde{\Theta}}_1\right)\!\!+\!\! |\mathbf{b}_1|^2\right)}{\sigma^2_1}\right)\nonumber\\
&\quad\quad\quad+c^{(n)}\left(\mathrm{Tr}\left(\mathbf{\tilde{\Theta}}_1\right)-\langle\partial||\mathbf{\tilde{\Theta}}_1||^{(n)}_2,\mathbf{\tilde{\Theta}}_1\rangle\right)\nonumber\\
&\quad\text{\text{s.t.}}\,\,\,\,\mathrm{C}_1: (1\!-\!\beta)\left(\mathrm{Tr}\left(\mathbf{R}_2\mathbf{\tilde{\Theta}}_1\right) \!+\! |\mathbf{b}_2|^2\right) \!\geq\! \left(2^{2\gamma_2}\!\!-\!\!1\right)\nonumber\\
&\quad\quad\quad\quad\quad\left((1\!-\!\beta)\left(\mathrm{Tr}\left(\mathbf{R}_1\mathbf{\tilde{\Theta}}_1\right)\! +\!|\mathbf{b}_1|^2\right)\!+\!\sigma^2_1\right),\nonumber\\
&\quad\quad\,\,\,\,\,\mathrm{C}_2: {(y^{(n)} \mathcal{X})^2+\left(\frac{\left(\mathrm{Tr}\left(\mathbf{R}_3\mathbf{\tilde{\Theta}}_1\right)\!+\!|b_3|^2\right)}{y^{(n)}}\right)^2}\nonumber\\
 &\quad\quad\quad\quad \leq2\mathrm{Tr}\left(\left(\mathbf{R}_4\mathbf{\tilde{\Theta}}_1\right)+|b_4|^2- \mathcal{X}\sigma^2_2\right)\nonumber\\
 &\quad\quad\,\,\,\,\,\mathrm{C}_3: \mathcal{X}+\beta\eta\left(\mathrm{Tr}\left(\mathbf{R}_5\mathbf{\tilde{\Theta}}_2\right) + |\mathbf{b}_5|^2\right)\times\nonumber\\
&\quad\quad\quad\,\,\,\,\,\left(\mathrm{Tr}\left(\left(\mathbf{R}_1+\mathbf{R}_2\right)\mathbf{\tilde{\Theta}}_1\right)\!+\! |\mathbf{b}_1|^2\!+\! |\mathbf{b}_2|^2\right) /\sigma^2_2\nonumber\\
&\quad\quad\quad\quad\quad\quad\quad\quad\quad\quad\quad\quad\quad\quad\quad \geq 2^{2\gamma_2}-1, \nonumber\\
&\quad\quad\,\,\,\,\,\mathrm{C}_4:{\left(\mathbf{\tilde{\Theta}}_1\right)_{m,m}=1}, m=1,2,...,M+1.\label{theta_111}
 \end{align}
then we can solve P5 by using standard convex optimization solvers such as CVX. $\mathbf{{\theta}}_1$ can be obtained by extracting the first $M$ rows of the last column of $\mathbf{\tilde{\Theta}}_1$. The proposed algorithm for optimizing $\mathbf{{\theta}}_1$ is summarized in Algorithm 1.

Finally, given $\beta$, $\mathbf{W}_1$, $\mathbf{W}_2$ and $\Theta_1$, $\mathbf{\theta}_2$ can be obtained by solving the following feasibility-check problem.
  \begin{align}
\text{\bf{P6}}:&{\mathrm{find} \quad\quad \tilde{\Theta}_2}\nonumber\\
&\quad\text{\text{s.t.}}\,\,\,\,\,\mathrm{C}_1:  \frac{\mathrm{Tr}\left(\mathbf{R}_4\mathbf{\tilde{\Theta}}_1\right) + |\mathbf{b}_4|^2}{\mathrm{Tr}\left(\mathbf{R}_3\mathbf{\tilde{\Theta}}_1\right) + |\mathbf{b}_3|^2+\sigma^2_2}+\beta\eta\times\nonumber\\
&\quad\quad\quad\quad\left(\mathrm{Tr}\left(\mathbf{R}_5\mathbf{\tilde{\Theta}}_2\right) \!\!+\!\! |\mathbf{b}_5|^2\right)\left(\mathrm{Tr}\left(\left(\mathbf{R}_1\!+\!\mathbf{R}_2\right)\mathbf{\tilde{\Theta}}_1\right)\right.\nonumber\\
&\quad\quad\quad\quad\quad\quad\quad\quad\left.+ |\mathbf{b}_1|^2+ |\mathbf{b}_2|^2\right)/\sigma^2_2 \geq 2^{2\gamma_2}-1, \nonumber\\
&\quad\quad\,\,\,\,\,\mathrm{C}_2:{{\mathrm{rank}\left(\mathbf{\tilde{\Theta}}_2\right)=1}}, \nonumber\\
&\quad\quad\,\,\,\,\,\mathrm{C}_3:{\left(\mathbf{\tilde{\Theta}}_2\right)_{m,m}=1}, m=1,2,...,M+1.\label{theta_2}
 \end{align}
The only challenge of this subproblem is the non-convex rank-one constraint, which can be transformed into the  equivalent form $\mathrm{Tr}\left(\mathbf{\tilde{\Theta}}_2\right)-||\mathbf{\tilde{\Theta}}_2||_2= 0$. Then, problem \text{\bf{P6}} can be solved by using the SCA-based method.

Based on the above analysis, we summarize the overall algorithm for jointly designing the beamforming of the AP, the phase reflection matrix of the RIS in phase 1 and phase 2, and the PS factor of $U_1$ in algorithm 2.
 \begin{algorithm}[htb]
 {{
 \caption{AO-based algorithm for P1 }
\begin{algorithmic}[1]
\STATE {\bf{Initialize:}} $\mathbf{{\theta}}^{(0)}_1$, $\mathbf{{\theta}}^{(0)}_2$;
\STATE{\bf{repeat}}
\STATE Update $\mathbf{w}^{(\!n\!)}_1, \mathbf{w}^{(\!n\!)}_2$ and $\beta^{(n)}$ from P2 with given $\mathbf{{\theta}}^{(n-1)}_1$, $\mathbf{{\theta}}^{(n\!-\!1)}_2$;
\STATE Update $\mathbf{{\theta}}^{(n)}_1$ according to algorithm 1 with given $\mathbf{w}^{(n)}_1$, $\mathbf{w}^{(n)}_2$, $\beta^{(n)}$ and $\mathbf{{\theta}}^{(n-1)}_2$;
\STATE Update $\mathbf{{\theta}}^{(n)}_2$ from P6 with given $\mathbf{w}^{(n)}_1$, $\mathbf{w}^{(n)}_2$, $\beta^{(n)}$ and $\mathbf{{\theta}}^{(n)}_1$;
\STATE $n-1 \rightarrow n$
\STATE {\bf{until}}\quad The objective function in P1 converges.
\end{algorithmic}
}}
 \end{algorithm}
\vspace{-0.2in}
\section{Numerical Results}
\begin{figure*}[htbp]
\centering
\subfigure[The feasible probability.]{
\begin{minipage}{0.3\linewidth}
\centering
\includegraphics[width=1\textwidth]{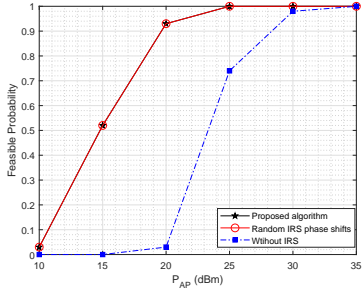}\\
\end{minipage}
}
\subfigure[The achievable rate of $U_1$.]{
\begin{minipage}{0.3\linewidth}
\centering
\includegraphics[width=1\textwidth]{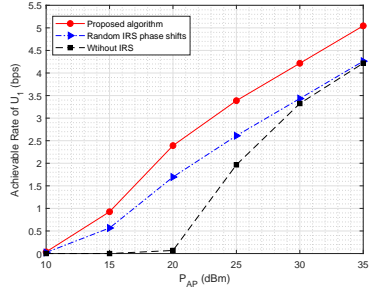}
\end{minipage}
}
\subfigure[Convergence behavior.]{
\begin{minipage}{0.3\linewidth}
\centering
\includegraphics[width=1\textwidth]{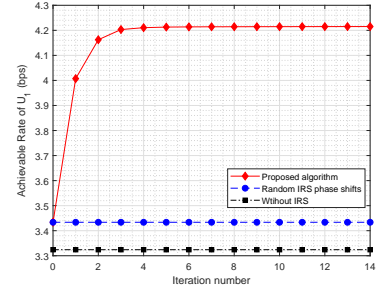}\\
\end{minipage}
}
\vspace{-0.1in}
\caption{Performance of the proposed algorithm}
\label{fig: C}
\vspace{-0.2in}
\end{figure*}

In this section, simulation results are provided to validate the effectiveness of the proposed algorithms. We assume that the AP is equipped with $N = 4$ transmit antennas and the number of the RIS reflection elements is $M=40$. It is assumed that the AP and RIS are located at $(0, 2, 0)$ m and $(11, 2, 0)$ m, respectively. The locations of $U_1$ and $U_2$ are $(8, 0, 0)$ m and $(12, 2, 0)$ m. The large-scale path loss is set as $PL = -30-10\alpha\log_{10}(d)$ dB, where $d$ is the transmission distance in meters and $\alpha$ represents the path loss exponent. Moreover, the pathloss exponents of both the HAP-$U_1$ and $U_1$-$U_2$ channels are set to $\alpha_1=3.5$, the HAP-$U_1$ channel is set to $\alpha_2=4$, and the RIS-assisted links are set to $\alpha_3=2$. The small-scale fading of all the RIS-assisted links follows a Rician distribution with Ricean factor 2, and all direct links follow Rayleigh fading \cite{guo2020weighted}. The noise variance at both users is set to $-50$ dBm and the target rate is $\gamma_2=0.5$ bit/s/Hz.

In order to evaluate the performance of the proposed algorithm, we consider the following two baselines: i) Random Phase: the initial values of $\mathbf{\theta}_1$ and $\mathbf{\theta}_2$ are randomly generated, and the other variables are optimized by using algorithm in \cite{xu2017joint}; ii) Without RIS: let $M = 0$, and then P1 is solved by the algorithm in \cite{xu2017joint}. Fig. \ref{fig: C}(a) and Fig. \ref{fig: C}(b) show the feasible probability and the achievable rate of $U_1$ of different schemes with respect to the transmit power of AP various from $10$ dBm to $35$ dBm. In Fig. \ref{fig: C}(a), it can be observed that the proposed scheme and the random phase scheme have the same performance in terms of the feasible probability, and they outperform the scheme without RIS. The scheme without RIS can achieve the same performance only if the SNR is very high. These observations confirm the ability of RIS to enhance the channel gain and improve the system performance. In Fig. \ref{fig: C}(b), one can see that the proposed scheme outperforms both baseline schemes over the whole power region in terms of the achievable rate of $U_1$. The scheme with random phase has an advantage over the scheme without RIS in the range of moderate SNR, while both of them have the same performance when the SNR is high. This is consistent with the trend of feasible probabilities in Fig. \ref{fig: C}(a).

Besides, the convergence behaviors versus the iteration number of the three schemes are shown in Fig. \ref{fig: C}(c). The transmit power of $P_\mathrm{AP}$ is set to 30 dBm. It can be observed that the proposed algorithm converges in about 4 iterations, validating its effectiveness.

\section{Conclusion}
\vspace{-0.09in}
In this work, we considered the performance optimization of an RIS-assisted C-NOMA with SWIPT network. Specifically, the RIS phase shifts in both transmission phases, the PS factor and the NOMA beamforming coefficients at the AP were jointly optimized to maximize the achievable rate of the strong user while guaranteeing the week user's QoS. The AO-based technique was applied to decompose the original problem and the complex phase shift constraints were addressed by using PBAGM algorithm. Numerical results demonstrated the effectiveness of the proposed algorithm.
\vspace{-0.1in}
  \begin{appendices}

\end{appendices}\renewcommand{\refname}{References}
\bibliography{Ref}
\bibliographystyle{myIEEEtran}

\end{document}